\documentclass[twocolumn]{aastex631}

\shortauthors{Mart\'inez-Mirav\'e, Tamborra, Vigna-G\'omez}
\graphicspath{{./}{figures/}}
\usepackage{amsmath}
\usepackage{amsfonts}
\usepackage{comment}
\usepackage{multirow}

\begin{document}

\title{Identifying Thorne-\.Zytkow Objects through Neutrinos}

\author[0000-0001-8649-0546]{Pablo Mart\'inez-Mirav\'e}
\affiliation{Niels Bohr International Academy and DARK, Niels Bohr Institute, University of Copenhagen, 
Blegdamsvej 17, 2100, Copenhagen, Denmark}

\author[0000-0001-7449-104X]{Irene Tamborra}
\affiliation{Niels Bohr International Academy and DARK, Niels Bohr Institute, University of Copenhagen, 
Blegdamsvej 17, 2100, Copenhagen, Denmark}

\author[0000-0003-1817-3586]{Alejandro Vigna-G\'omez}
\affiliation{Max Planck Institute for Astrophysics, Karl-Schwarzschild-Strasse 1, 85748, Garching bei München, Germany}

\begin{abstract}
Thorne-{\.Z}ytkow Objects (T$\dot{\rm Z}$Os) have been predicted to form when a neutron star is engulfed by a diffuse, convective giant envelope. Accretion onto a neutron star at a rate that is larger than $10^{-4}\, M_\odot$~yr$^{-1}$ is expected to lead to significant emission of neutrinos of all flavors with energy of $1$--$100$~MeV. Since the neutrino signal is expected to largely vary in time (from milliseconds to thousands of years), we outline detection strategies tailored to the signal duration. We find that neutrino detection from T$\dot{\rm Z}$Os up to the Small Magellanic Cloud is within the reach of current- and next-generation neutrino observatories, such as Super- and Hyper-Kamiokande, the IceCube Neutrino Observatory, and JUNO. Interestingly, if targeted searches for neutrinos from T$\dot{\rm Z}$O candidates (e.g.~VX Sgr in our Galaxy as well as HV 2112 and HV 11417 in the Small Magellanic Cloud) should lead to positive results, neutrinos could positively identify the nature of such sources and their accretion rate. Furthermore, the diffuse supernova neutrino background may be able to rule out extreme scenarios for the formation and accretion rates of T\.ZOs. Our findings should serve as motivation for establishing dedicated searches for neutrino emission from T$\dot{\rm Z}$Os. 
This is especially timely since it is challenging to detect T\.ZOs via electromagnetic radiation unambiguously,
and the T$\dot{\rm Z}$O gravitational wave signal could be probed with next-generation detectors for sources within our Galaxy only. 
\end{abstract}

\section{Introduction} \label{sec:intro}

Thorne-\.Zytkow Objects (T$\dot{\rm Z}$Os) are stellar bodies characterized by a compact object surrounded by an extended and diffuse envelope~\citep{OGrady:2024ioe,Grichener:2024mmg}. T$\dot{\rm Z}$Os are  expected to form  when a neutron star  is engulfed by another star, likely a red giant or supergiant~\citep{Thorne:1975,Thorne:1977}. 
Potential formation channels of T$\dot{\rm Z}$Os include binary coalescence, following  a common-envelope phase, and  direct collision of the  two stellar objects  in dense dynamical environments. 

As the neutron star pierces through the outer layers of the giant companion, the stellar  envelope  accretes onto the neutron star. During the inspiral phase, the core of the red supergiant spins up~\citep{Hutchinson-Smith:2023apu}. If the angular momentum of postmerger material is larger than the one needed to orbit at the innermost stable circular orbit around the neutron star, a disk forms.
Several electromagnetic transients could stem from this process, such as  long gamma-ray burst~\citep{Zhang:2000ak,Levan:2013gcz}, luminous fast blue optical transients~\citep{Soker:2018msh,Metzger:2022xep}, common envelope jets supernovae~\citep{Grichener:2018way}, and merger-driven core-collapse supernovae~\citep{Fryer:1996}. This class of transient sources could host  rapid neutron capture ($r$) process nucleosynthesis~\citep{Papish:2015}.

Alternatively, if at the end of the inspiral phase the gas surrounding the compact object has acquired has an angular momentum lower than that required to form a disk, accretion onto the neutron star is quasi-spherical, leading to the formation of a classical T\.ZO, as originally envisioned \citep{Thorne:1975, Thorne:1977}. These stable T$\dot{\rm Z}$Os have been considered as production sites of heavy proton-rich isotopes through the rapid proton capture ($rp$) process~\citep{1993MNRAS.263..817C, Nava-Callejas:2024rkv}.

In both cases, accretion onto the central compact core occurs at an hypercritical (super-Eddington) rate,  resulting in high-density and temperature regions (\mbox{$\rho \gtrsim 10 ^ 5$ g cm$^{-3}$} and T $\gtrsim 10 ^{10}$ K) in the surroundings of the central compact object.
If the mass of the latter exceeds the Tolman–Oppenheimer–Volkoff limit, a black hole forms. 

\begin{figure}[ht]
\includegraphics[width = \linewidth]{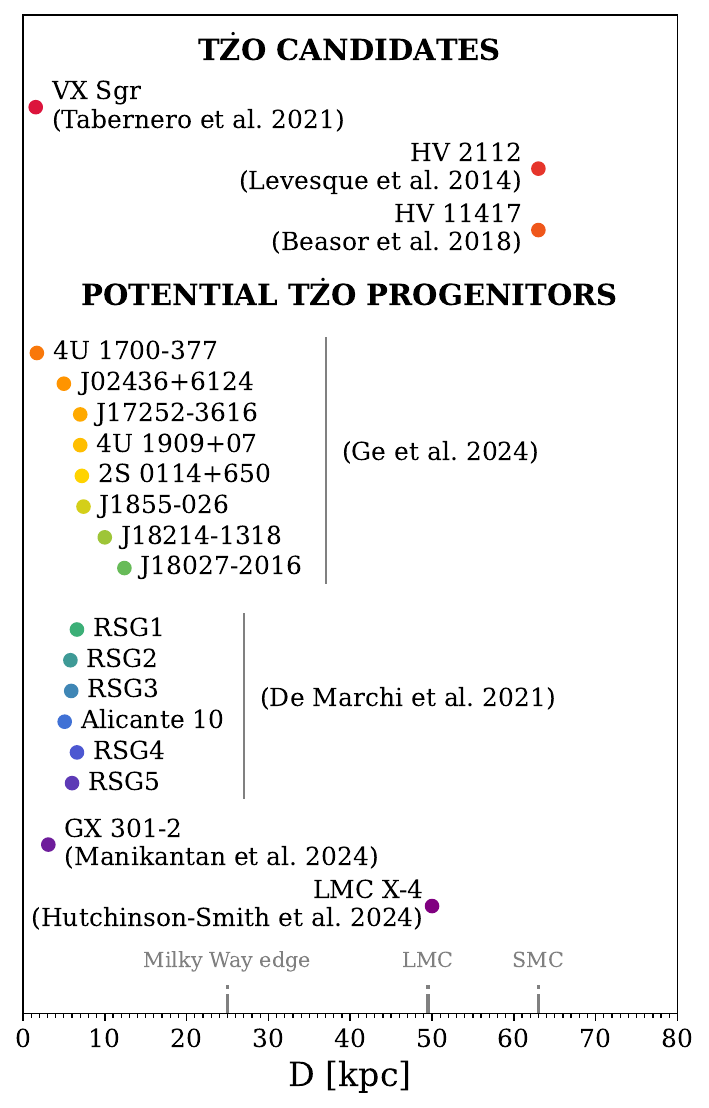}
\caption{T$\dot{\rm Z}$O candidates~\citep{Tabernero:2021,Levesque:2014,Beasor:2018} and potential T$\dot{\rm Z}$O progenitors in the Galaxy, as well as in the Large Magellanic Cloud (LMC) and in the Small Magellanic Cloud (SMC). Potential candidates are considered to be  (i) high-mass X-ray binaries in which the neutron star accretes matter from the stellar companion and unstable mass transfer is expected~\citep{Ge:2024key,Manikantan:2023azm, Hutchinson-Smith:2023apu} and (ii) stellar clusters where red supergiants are abundant~\citep{DeMarchi:2021vwr}. 
\label{fig:sources}}
\end{figure}

Population synthesis models predict $\mathcal{O}(1$--$100)$ T$\dot{\rm Z}$Os~in the Milky Way~\citep{1995MNRAS.274..485P,Nathaniel:2024mde}. 
Figure~\ref{fig:sources} displays a collection of candidates  monitored in the Local Group~\citep{Kuchner:2002,Levesque:2014,Beasor:2018,Tabernero:2021,Ogrady:2023}. 
In addition, as illustrated in Fig.~\ref{fig:sources}, nearby high-mass X-ray binaries (HMXBs) may  undergo unstable mass transfer~\citep{Hutchinson-Smith:2023apu,Ge:2024key,Manikantan:2023azm}, potentially leading to the formation of  T$\dot{\rm Z}$Os. Potential  T$\dot{\rm Z}$O progenitors can also be found in stellar clusters in the Milky Way, which are  rich in red supergiants~\citep{DeMarchi:2021vwr}.  

Observationally, T$\dot{\rm Z}$Os are expected to resemble red supergiants, or nitrogen-rich Wolf–Rayet stars (if the star is not stripped of the hydrogen layer), with peculiar spectroscopic lines  coming from the nucleosynthesis occurring in the source~\citep{Biehle:1994,Podsiadlowski:1995,Farmer:2023rur,Nava-Callejas:2024rkv}. Since it is challenging to assess the presence of a neutron-star core solely relying on the electromagnetic signatures, a multi-messenger approach involving the detection of gravitational waves from the rotating neutron star or a common-envelope evolution phase has been invoked to allow us for the unambiguous identification of  T$\dot{\rm Z}$Os~\citep{DeMarchi:2021vwr,Nazin:1995bp,DeMarchi:2021vwr,Renzo:2021aho,Moran-Fraile:2023tvg}. 

In this {\it Letter}, we propose  neutrinos as key messengers to identify these fascinating stellar objects, even with neutrino observatories currently taking data. In fact, for  accretion rates ranging from $10^{-4}\, M_\odot$~yr$^{-1}$ to $10^{6}\, M_\odot$~yr$^{-1}$, we estimate that the temperature in the vicinity of the neutron star is $\sim \mathcal{O}(1$--$10)$~MeV. Such conditions could lead to the copious production of neutrinos with energy in the $10$--$100$~MeV range~\footnote{Notice that for T$\dot{\rm Z}$Os with  significantly lower accretion rates, e.g.~the case of $10^{-8}\, M_\odot$~yr$^{-1}$ discussed in~\citealt{Cannon:1992}, neutrino production is negligible.}. 

We compute the neutrino emission from T$\dot{\rm Z}$Os with accretion rates larger than $10^{-4}$ $M_\odot$~yr$^{-1}$. We tailor the detection strategy according to the duration of the expected neutrino emission.
The inclusion of neutrinos in T$\dot{\rm Z}$O searches is especially timely, given that current gravitational wave detectors are not sensitive to  these objects.
We introduce the source models in \S~\ref{sec:source} and the thermal neutrino spectral distribution  expected from  T$\dot{\rm Z}$Os in \S~\ref{sec:nu-emission}.
We then investigate the detection prospects of neutrinos from  T$\dot{\rm Z}$Os in \S~\ref{sec:nu-detection} in existing  and next-generation neutrino observatories and assess the  contribution of T$\dot{\rm Z}$Os to the diffuse background of astrophysical neutrinos in \S~\ref{sec:nu-diffuse}. We discuss and summarize our findings in \S~\ref{sec:conclusions}. Additional details on the T$\dot{\rm Z}$O source models and the neutrino production rates are presented in Appendix~\ref{sec:app-source-nu}, while  Appendix~\ref{sec:app-nu-obs} focuses on the modeling of the expected event rates in neutrino observatories, respectively.

\section{Source model}
\label{sec:source}

We focus on T$\dot{\rm Z}$Os with hypercritical accretion rates from $10^{-4}\, M_\odot$~yr$^{-1}$ to $10^{6}\, M_\odot$~yr$^{-1}$ (i.e. $3\times 10^{-12}$ to $3\times 10^{-2}\, M_\odot$ s$^{-1}$, which is $3$--$13$ orders of magnitude larger than the critical, Eddington-limited rate), since for such cases neutrinos are expected to be the primary energy loss channel. 
For $\dot{M} \simeq 10^{-4}\, M_\odot$~yr$^{-1}$, the photon trapping radius is several orders of magnitude larger than the neutron-star radius ($r_{\rm trappping} \sim 10^4$ km  and grows linearly with the accretion rate, see~\citealt{Houck:1991a}). Hence, neutrino losses are the main cooling mechanism.

While neutrino emission takes place already in the inspiraling phase~\citep{Hutchinson-Smith:2023apu}, in this work we consider steady models focusing on the regions surrounding the proto-neutron star where neutrinos are expected to be most copiously produced.
As detailed in Appendix~\ref{sec:app-source-nu}, for accretion rates $10^{4}\, M_\odot$~yr$^{-1}\leq  \dot{M } \leq 10^{6}\, M_\odot$~yr$^{-1}$, we assume that an accretion disk forms around the neutron star when the inspiraling ends. We model the density and temperature profiles following \cite{DiMatteo:2002iex,Zhang:2007ds}. This model assumes steady accretion onto a compact object such that the outer region is advection dominated and the inner region is neutrino dominated. Furthermore, for radius smaller than $\sim$20 km, the model adopts a self-similar solution for the profiles to account for the fact that the compact object is a neutron star. We assume that the disk has a radius of $1000$~km, i.e.~it is $100$ times larger than the size of the neutron star. In the surroundings of the neutron star, the temperature  is $T \sim 4\times 10^{10}$--$8\times 10^{10}$~K and the baryon density is $\rho \sim 10^9$--$10^{11}$~g~cm$^{-3}$.

For accretion rates in the range $10^{-4}\, M_\odot$~yr$^{-1}\leq  \dot{M }< 10^{4}\, M_\odot$~yr$^{-1}$,  we compute the neutrino emission rates relying on a steady spherically-symmetric source, building on the model presented in~\cite{Chevalier:1989}. This model assumes steady accretion of shocked gas from the envelope onto the neutron star, subsonic  velocity of the shocked matter, and that the postshock flow is adiabatic. As a result, the medium temperature and baryon density profiles scale as power laws as functions of the radius in the proximity of the neutron star (i.e., T$(r) \sim r^{-1}$ and $\rho(r)\sim r^{-3}$) with characteristic temperature and baryon density of $T \sim 10^{10}$--$10^{11}$~K and $\rho \sim 10^5$--$10^9$~g~cm$^{-3}$. In both scenarios, we consider a neutron star with a mass of $1.4 \,M_\odot$ and  radius of $10$~km (but the density and temperature profiles depend mildly on these choices).

\section{Neutrino emission}
\label{sec:nu-emission}

For a neutron star of mass $M_{\rm NS}$ and radius $r_{\rm NS}$, which accretes a mass $\Delta m$ at a constant rate, we assume that the energy radiated in neutrinos is a fraction of the gain in gravitational energy of the system. Hence, we define the upper limit on the duration of the neutrino signal as the time to black hole formation:
\begin{align}
    \tau \simeq \frac{1}{L_\nu} \frac{G {M}_{\rm NS} \Delta m}{{r}_{\rm NS}} \, ,
    \label{eqn:tbh}
\end{align}
where $G$ is the gravitational constant and $L_{\nu}$ is the neutrino luminosity for all flavors.
From Eq.~\ref{eqn:tbh}, we deduce that the duration of the neutrino signal depends on the amount of accreted mass. Note,  that $\tau$ provides an upper limit on the actual duration of the neutrino emission, since (i) the amount of accreted mass before collapse might be smaller than $\Delta m$ and (ii) accretion might not be constant  for extended periods of time. Hence, hereafter, we denote the duration of the neutrino emission with $\Delta t_\nu \leq \tau$.

If accretion onto the neutron star occurs at a fast rate (i.e., $\dot{M} \geq 10^4 M_\odot$~yr$^{-1} \approx 10^{-12} M_\odot$~s$^{-1}$), the formation of a T$\dot{\rm Z}$O~would be halted by the collapse of the neutron star into a black hole~\citep{Chevalier:1993,Fryer:1996}. In such cases, neutrino emission would cease promptly (hereafter named ``T$\dot{\rm Z}$O burst'' scenario). We identify this transient scenario with the one in which we expect the formation of an accretion disk. Otherwise, for lower accretion rates, $10^{-4} M_\odot$~yr$^{-1} \leq\dot{M} < 10^4 M_\odot$~yr$^{-1}$ we expect a steady neutrino emission (``steady T$\dot{\rm Z}$O''  scenario) arising from steady spherically-symmetric accretion. Table~\ref{tab:evrates} summarizes the upper limit on the duration of the neutrino signal expected for different accretion rates, assuming $\Delta m = 1 M_\odot$.

In the innermost regions of the accretion disk, the large densities and temperatures can favor helium photodissociation  and beta process ($e^- + p \rightleftharpoons \nu_e + n$ and $e^+ + n \leftrightharpoons \bar{\nu}_e + p$).  For accretion rates $\dot{M} > 10^4 M_\odot$~yr$^{-1}$, these processes dominate the neutrino production in the disk, as detailed in  Appendix~\ref{sec:app-source-nu}.
Conversely, in the region surrounding the spherically-accreting central neutron star and extending tens of kilometers outwards, the dominant neutrino emission channel is electron-positron pair annihilation (cf. \cite{Esteban:2023uvh} for comparable findings), as illustrated in Appendix~\ref{sec:app-source-nu} (cf.~Fig.~\ref{fig:elr}).
In the core of the neutron star, due to the large density, neutrinos are mainly produced thanks to plasmon decay, but are trapped.
Other neutrino emission channels such as bremsstrahlung and photoneutrino production negligibly contribute  to the expected neutrino flux from T$\dot{\rm Z}$Os, as shown in Fig.~\ref{fig:elr}.

\begin{figure}
    \centering
    \includegraphics[width=\linewidth]{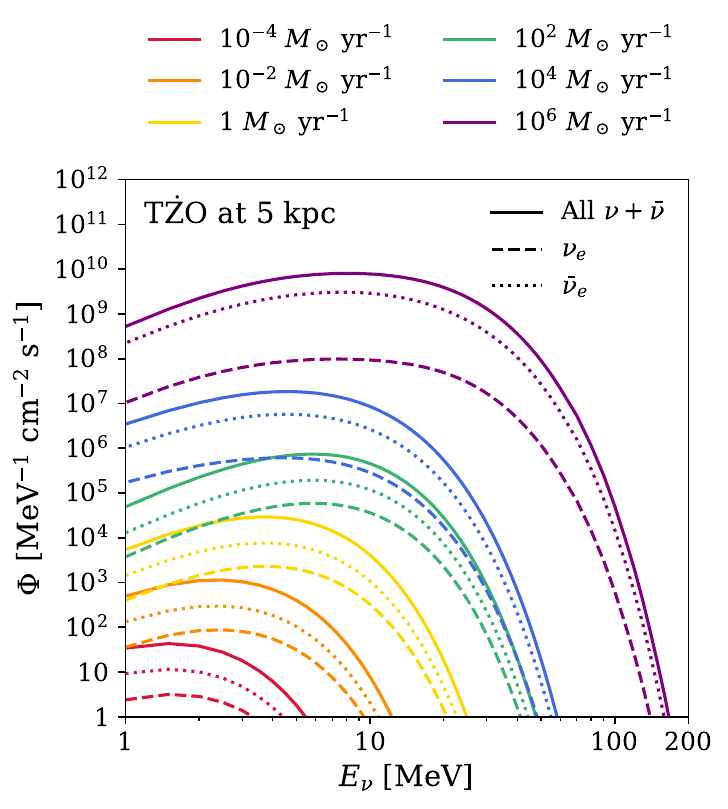}
    \caption{Neutrino flux  from a T$\dot{\rm Z}$O at  $5$~kpc from Earth, taking into account flavor conversion. The color lines correspond to the accretion rates considered in this work: $\dot{M} = 10^{-4}\,{M}_\odot$~yr$^{-1},\,10^{-2}\,{M}_\odot$~yr$^{-1},\, 1\,{M}_\odot$~yr$^{-1},\,$ \mbox{$10^{2}\,{M}_\odot$~yr$^{-1},$}$\, 10^{4}\,{M}_\odot$ yr$^{-1}$, and $10^{6}\,{M}_\odot$~yr$^{-1}$, from bottom to top. Dashed, dotted, and solid lines correspond to electron neutrinos, electron antineutrinos, and the total flux of all neutrinos and antineutrinos, respectively. The neutrino flux  increases as a function of the  accretion rate, peaking at  higher energies. \label{fig:nuflux}}
    \end{figure}
Figure~\ref{fig:nuflux} displays the neutrino flux for all flavors  from a T$\dot{\rm Z}$O at  $5$~kpc from Earth. For large accretion rates, the temperature in the innermost regions increases (cf.~Fig.~\ref{fig:elr}) and neutrino emission through beta process is significantly enhanced, overcoming the emission from pair annihilation. As a result, the expected neutrino flux is larger, the peak of the spectrum  shifts to higher energies and small differences in the spectral shape are found at the lowest energies. Note that electron neutrinos and antineutrinos are produced via charged current and neutral current interactions, hence the electron flavors in the production region are more abundant than the non-electron ones (see also dotted and dashed lines in Fig.~\ref{fig:nuflux} for $\bar\nu_e$ and $\nu_e$, respectively).
We stress that, since we rely on a steady T$\dot{\rm Z}$O source model, the neutrino emission that we predict is constant over time. In this sense, we provide an upper limit of the expected neutrino flux.

\section{Neutrino detection prospects}
\label{sec:nu-detection}

\begin{deluxetable*}{ccccccc}[tpb!]
    \tablecolumns{2} \tablecaption{Upper limit on the duration of the neutrino signal $\tau$ (for $\Delta m = 1 M_\odot$, see Eq.~\ref{eqn:tbh}) and expected     neutrino event rates at IceCube, JUNO, Super-Kamiokande, and Hyper-Kamiokande, respectively, from a T$\dot{\rm Z}$O at $5$~kpc from Earth and for different accretion rates.}
    \label{tab:evrates}
    \tablewidth{0pt}
    \tablehead{ & \colhead{$\dot{M}$} & \colhead{$\tau$} & \colhead{IceCube} & \colhead{Super-Kamiokande}  & \colhead{JUNO}  & \colhead{Hyper-Kamiokande} \\ & \colhead{[$M_\odot$ yr$^{-1}$]} & [s] &  Signal rate [s$^{-1}$] & Signal rate [s$^{-1}$]  & Signal rate [s$^{-1}$] & Signal rate [s$^{-1}$]}
    \startdata
    \multirow{4}{*}{Steady T$\dot{\rm Z}$O} & $10^{-4}$ & $1.8\times 10^{11}$ & - & $\sim \mathcal{O}(10^{-10})$ & $\sim \mathcal{O}(10^{-11})$ &$\sim \mathcal{O}( 10^{-9})$\\
    &  $10^{-2}$ & $2.7\times 10^9$ & - & $1.2\times10^{-7}$ & $5.7\times 10^{-8}$ & $1.6\times 10^{-6}$\\
    & $1$ & $4.3\times 10^7$ &- & $5.0 \times 10^{-5}$ & $4.9\times 10^{-5}$ &$6.4\times 10^{-4}$\\
    & $10^{2}$ & $6.7\times 10^5$ &- & $8.4\times10^{-3}$  &$1.2\times10^{-2}$ &$1.1\times10^{-1}$\\ \hline
   \multirow{2}{*}{T$\dot{\rm Z}$O burst}& $10^{4}$ & $1.1 \times 10^4$& $8.0\times 10^2$ & 14 & - & $1.1\times 10^2$ \\
     & $10^{6}$ & $2.0\times 10^2$ & $9.6\times 10^6$& $6.4\times 10^3$ & - & $5.3\times 10^{4}$\\
    \enddata
    \end{deluxetable*}
    
In the following, we explore the detection prospects of neutrinos from T$\dot{\rm Z}$Os in existing and upcoming  neutrino observatories. We consider the IceCube Neutrino Observatory, Super-Kamiokande and Hyper-Kamiokande, as well as JUNO. Because of  the large variation of the expected duration of the neutrino signal as a function of the accretion rate, we propose tailored detection strategies to enhance the expected sensitivity and distinguish between T$\dot{\rm Z}$O bursts (i.e., signal duration $\Delta{t}_{\nu} \gtrsim 100$~s and accretion rates $\dot{M} \geq 10^{4} M_\odot$~yr$^{-1}$) and steady T$\dot{\rm Z}$Os  (i.e., signal duration of months to years and accretion rates $10^{-4} M_\odot$~yr$^{-1} \leq \dot{M} \leq 10^{4} M_\odot$~yr$^{-1}$). 

\subsection{T$\dot{\rm Z}$O bursts}

In order to investigate the neutrino detection prospects, we consider  operative water-Cherenkov neutrino detectors, such as the IceCube Neutrino Observatory~\citep{IceCube:2011cwc} and Super-Kamiokande~\citep{Super-Kamiokande:2016kji,Super-Kamiokande:2024pmv}, as well as the next-generation  Hyper-Kamiokande~\citep{Hyper-Kamiokande:2018ofw} which is expected to have increased sensitivity with respect to Super-Kamiokande due to the larger fiducial volume. As detailed in Appendix~\ref{sec:app-nu-obs},  neutrinos are mainly detected via inverse beta decay (IBD, $\bar{\nu}_e + p \to n + e^+$). 
However, quasi-elastic scattering of electron (anti)neutrinos  on oxygen ($\nu_e + ^{16}{\rm O} \to e^- + X$ and $\bar{\nu}_e + ^{16}{\rm O} \to e^+ + X$, where $X$ denotes the final state nucleus) and elastic scattering of all flavor (anti)neutrinos  on electrons ($\nu_\alpha + e^- \to \nu_\alpha + e^-$ and $\bar{\nu}_\alpha + e^- \to \bar{\nu}_\alpha + e^-$, for each flavor $\alpha$) also contribute to the signal event rate. 

\begin{figure*}
    \centering
\includegraphics[width=0.49\linewidth]{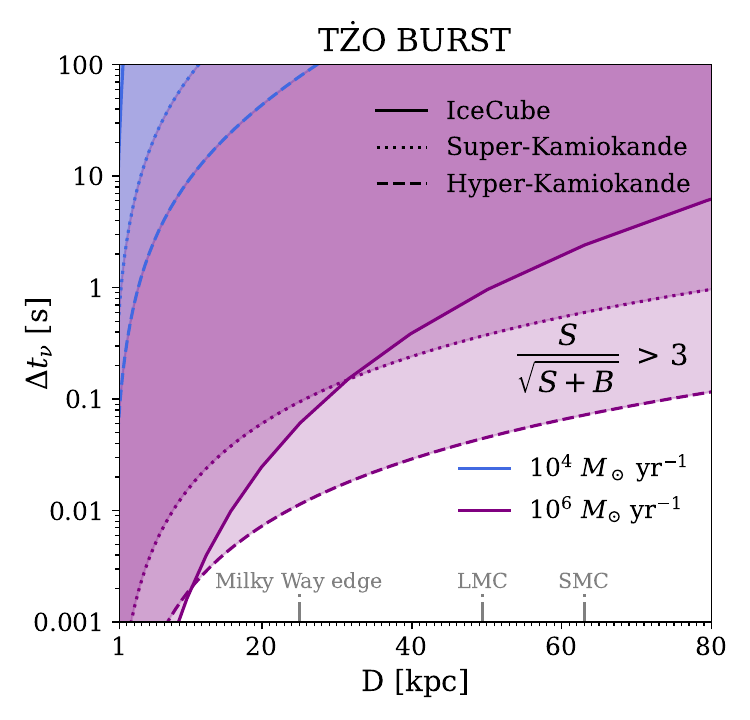}
    \includegraphics[width =0.49\linewidth]{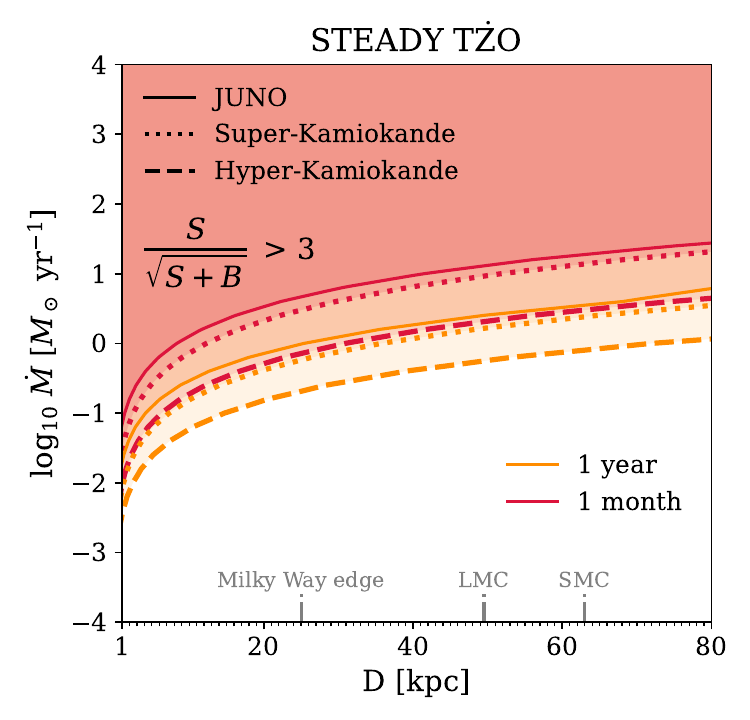}
    \caption{Detection prospects of neutrinos from T$\dot{\rm Z}$Os. {\it Left panel:} Sensitivity of neutrino detection from  T$\dot{\rm Z}$O  bursts detected by IceCube (solid lines), Super-Kamiokande (dotted lines) and Hyper-Kamiokande (dashed lines) in the plane spanned by the source distance from Earth  and the signal duration. In order to guide the eye, the Milky Way edge is marked on the x-axis, together with the Large and Small Magellanic Cloud (LMC and SMC, respectively). The shaded regions correspond to a signal-to-noise ratio larger than $3$,  for constant accretion rates of $10^{4}\, M_\odot$~yr$^{-1}$ and $10^6\, M_\odot$~yr$^{-1}$, in blue and purple, respectively. 
    Super-Kamiokande (Hyper-Kamiokande) and IceCube are sensitive to  neutrinos bursts from T$\dot{\rm Z}$Os, lasting for  $\gtrsim \mathcal{O}(0.1)$~s,    within our Galaxy and beyond the Small Magellanic Cloud.
    {\it Right panel:} Same as left panel, but for steady T$\dot{\rm Z}$Os detected by Super-Kamiokande (dotted lines), JUNO (solid lines), and Hyper-Kamiokande (dashed lines).
    The orange and red bands correspond to  data-taking windows of one year and one month, respectively. 
    Super-Kamiokande and JUNO should be sensitive to the neutrino emission from steady T$\dot{\rm Z}$Os beyond the Small Magellanic Cloud for $\dot{M} \gtrsim 0.1\, M_\odot$~yr$^{-1}$. 
    \label{fig:prospects}}
\end{figure*}

Figure~\ref{fig:prospects}  (left panel) displays the sensitivity of IceCube, Super-Kamiokande, and Hyper-Kamiokande to detect neutrinos from T$\dot{\rm Z}$O bursts at a distance $D$ from Earth and lasting for $\Delta t_{\nu}$ (note that the difference in the event rate between Super-Kamiokande and Hyper-Kamiokande is mainly a scaling factor due to the roughly ten times larger effective volume of Hyper-Kamiokande). 
In Fig.~\ref{fig:prospects}, for $\dot{M} = 10^4 M_\odot$~yr$^{-1}$ and $10^6  M_\odot$~yr$^{-1}$, we highlight in color the regions of the plane spanned by the T$\dot{\rm Z}$O distance from Earth and $\Delta t_{\nu}$  where the signal-to-noise ratio would be larger than $3$ (${S}/\sqrt{S+B}> 3$, where $S$ and $B$ denote the number of signal and background events, respectively); this  corresponds to the $3 \sigma$ exclusion of the signal hypothesis in the absence of additional systematic uncertainties.

IceCube has a larger effective area  than Super-Kamiokande (and Hyper-Kamiokande), therefore it can collect a larger number of signal events. Nevertheless, it  features a significant rate of background events \citep[$1.5\times10^6$ s$^{-1}$]{IceCube:2011cwc}, with respect to Super-Kamiokande, which has negligible backgrounds for burst-like searches. As a consequence, Super-Kamiokande has a better sensitivity to T$\dot{\rm Z}$Os  at larger distances (cf.~Fig.~\ref{fig:prospects} and Table~\ref{tab:evrates}) than IceCube. Moreover, Super-Kamiokande allows to reconstruct  spectral energy information of the detected neutrinos, not available at IceCube. 

The left panel of Fig.~\ref{fig:prospects} shows that, for T$\dot{\rm Z}$O bursts with accretion rate equal to $10^6 M_\odot$~yr$^{-1}$ ($10^4 M_\odot$~yr$^{-1}$), we would be able to detect neutrinos up to the Small Magellanic Cloud for $\Delta t_{\nu} \gtrsim 0.1$~s (the Milky Way edge for $\Delta t_{\nu} \gtrsim 100$~s).
Such detection prospects clearly illustrate the potential of existing neutrino observatories of monitoring T$\dot{\rm Z}$Os in a larger fraction of the local Universe than future gravitational waves detectors would be able to do. In fact, the sensitivity of the latter is expected to be limited to Galactic events~\citep{DeMarchi:2021vwr,Renzo:2021aho}.

\subsection{Steady T$\dot{\rm Z}$Os}
If accretion is sustained for relatively long periods of time (i.e., at least several hours), a different detection strategy at neutrino observatories should be employed. In fact, to capture a steady neutrino signal lasting for months to years, we should search for an excess of neutrino events over background for a given time window. 
To this purpose, the neutrino telescopes Super-Kamiokande, JUNO~\citep{JUNO:2015zny}, and Hyper-Kamiokande would lead the detection efforts. 

We focus on the potential of Super-Kamiokande  loaded with Gadolinium to enhance its sensitivity to electron antineutrinos~\citep{Super-Kamiokande:2021the,Super-Kamiokande:2024kcb}. Our projections assume that Hyper-Kamiokande would have a similar setup, while being larger in size. 
By adopting the same region of interest  used for searches of the diffuse supernova neutrino background (DSNB),  we could look for an electron antineutrino flux from T$\dot{\rm Z}$Os through inverse beta decay in the energy window where the backgrounds are under control. Hence,  we consider Super-Kamiokande and Hyper-Kamiokande with   a detection efficiency of $55\%$ (corresponding to water loaded with Gadolinium) in the energy window between $9.3$  and $31.3$~MeV, with a background rate of \mbox{$1.0\times10^{-6} $~s$^{-1}/(100$~kT)}~\citep{Super-Kamiokande:2021jaq}.  As for JUNO, we assume a detection efficiency of $80\%$ for energies between $12$ and $30$~MeV, with a background event rate of \mbox{$1.4\times 10^{-7} $~s$^{-1}/(17$~kT)}~\citep{JUNO:2022lpc}. We refer the reader to Appendix~\ref{sec:app-nu-obs} for additional details.

Figure~\ref{fig:prospects} (right panel) displays  the sensitivity of Super-Kamiokande, Hyper-Kamionade, and JUNO in the plane spanned by the  mass accretion rate and the distance of the source from Earth for two different  time windows of data taking of one month (in red) and one year (in orange), respectively. As for the case of T$\dot{\rm Z}$O bursts in the left panel of the same figure, we highlight in color the regions of the parameter space with a signal-to-noise ratio larger than $3$. 
We can see that neutrinos from T$\dot{\rm Z}$Os with accretion rates larger than $0.1\, M_\odot$~yr$^{-1}$ can be  detected at Super-Kamiokande and JUNO, even if the source is located beyond the Small Magellanic Cloud.  Table~\ref{tab:evrates} also summarizes the signal event rates expected for steady T$\dot{\rm Z}$Os.

\subsection{Can we rely on neutrino detection to identify  T\.ZOs?}
Figure~\ref{fig:sources} indicates that, among the T$\dot{\rm Z}$O candidates, VX Sgr~\citep{Tabernero:2021} is at $\sim 1.5$--$1.7$~kpc. If its accretion rate is $\gtrsim 10^{-2}\, M_\odot$~yr$^{-1}$,   neutrino searches  guided by Super-Kamiokande, IceCube, and soon by JUNO would be able to identify this  source  as a  T$\dot{\rm Z}$O, since large neutrino event  statistics should be observed, independent of the  theoretical uncertainties on the source properties. On the other hand, if dedicated searches should  not lead to any positive neutrino detection, this may be an indication that VX Sgr is not a T$\dot{\rm Z}$O or its accretion rate is so low that neutrino production is inefficient.

As for HV 2112~\citep{Levesque:2014} and HV 11417~\citep{Beasor:2018}, these sources are at  $~\sim 60$~kpc from Earth. We could adopt Super-Kamiokande and IceCube to test whether these sources  have an accretion rate around $10^6\, M_\odot$~yr$^{-1}$ or larger. If these sources should be steady T$\dot{\rm Z}$Os, then one year of neutrino data taking with Super-Kamiokande (JUNO) would constrain accretion rates above $2.4\, M_\odot$~yr$^{-1}$ ($3.2\, M_\odot$~yr$^{-1}$). 

As for the potential T$\dot{\rm Z}$O progenitors shown in Fig.~\ref{fig:sources}, the vast majority of such sources~\citep{Ge:2024key,DeMarchi:2021vwr,Manikantan:2023azm} is located within $10$~kpc from Earth. This implies that, if these or similar sources were to become steady T$\dot{\rm Z}$Os, we should expect to detect a cumulative flux of neutrinos with existing (or soon-to-be operative) detectors for accretion rates $\gtrsim 10^{-2}\, M_\odot$~yr$^{-1}$ after one year of data taking. If, instead, these candidates should evolve into T$\dot{\rm Z}$O bursts, then we would detect their neutrino emission for $\dot{M} \gtrsim 10^5 M_\odot$~yr$^{-1}$.

There may be additional undetected candidates within the Local Group that could undergo a merger in the near future. 
Most of our current understanding of such events comes from low-mass stellar mergers in the Galaxy, such as the archetypal luminous red nova V1309 Sco~\citep{2011A&A...528A.114T}.
However, for the more massive T$\dot{\rm Z}$O progenitors, the merger has been predicted to result in a luminous merger-driven explosion~\citep{2020ApJ...892...13S}.
One promising prospect for the electromagnetic detection of such mergers is, for example, the Rubin Observatory/LSST~\citep{2009arXiv0912.0201L}.
A coincident detection of a luminous explosion with neutrino emission  could provide compelling evidence of the formation of T$\dot{\rm Z}$Os.
In addition, it can shed light on the physics behind the collapse of a neutron star into a black hole, as well as on the formation of T$\dot{\rm Z}$Os with a low-mass thin-envelope~\citep{2024ApJ...971..132E}.

Given that Super-Kamiokande is currently taking data and JUNO is expected to become operative within the next year, we urge to carry out searches of neutrinos from T$\dot{\rm Z}$Os. Such targeted programs, for burst-like and steady T$\dot{\rm Z}$Os, will be crucial  to assess the nature of existing T$\dot{\rm Z}$O candidates, eventually providing crucial insight on the nature of the source core that the electromagnetic data cannot probe. This is especially timely in light of the fact that any gravitational wave signal from T$\dot{\rm Z}$Os could be detected only with next-generation detectors~\citep{DeMarchi:2021vwr,Renzo:2021aho}.

\section{Diffuse neutrino flux from T$\dot{\rm Z}$Os}
\label{sec:nu-diffuse}

\begin{figure}[t!]
    \centering
    \includegraphics[width=\linewidth]{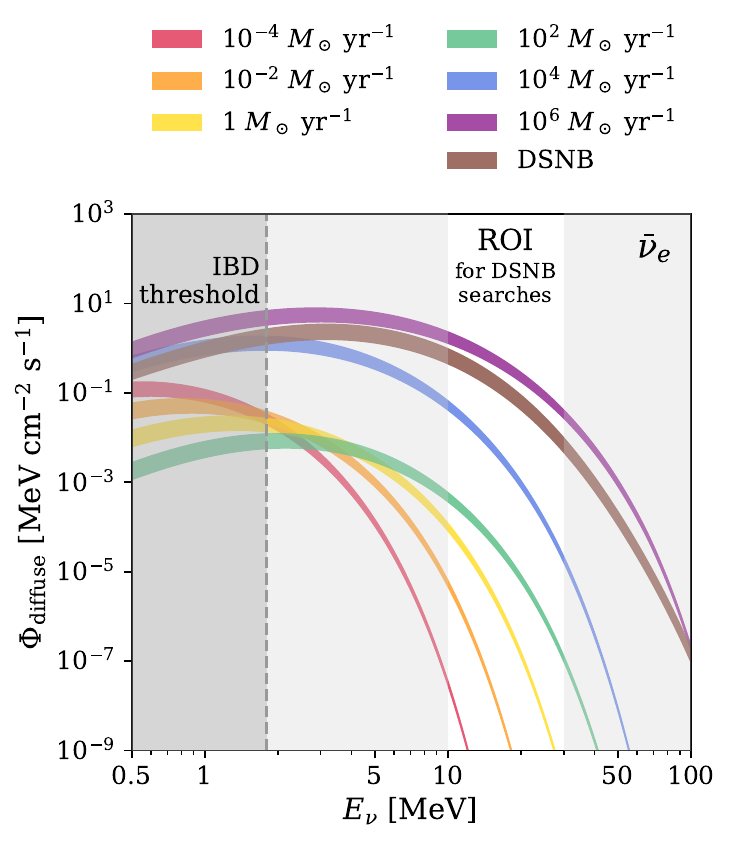}
    \caption{Diffuse flux of electron antineutrinos from T$\dot{\rm Z}$Os as a function of the neutrino energy, assuming that all T$\dot{\rm Z}$Os accrete at the same rate.  The colored bands correspond to the accretion rates considered in this work, and the width of each band represents the uncertainty on the local rate of T$\dot{\rm Z}$O (see Eq.~\ref{eq:rate}). The region of interest (ROI) for detection  coincides with the  one from the DSNB seaches (i.e., the white vertical band between $10$ and $30$~MeV). The energy threshold for inverse beta decay (IBD, the most promising detection channel) is marked by the dashed line at $1.8$~MeV. For comparison, the brown band represents  the  DSNB computed as in~\cite{Martinez-Mirave:2024zck}. Current searches for the DSNB already exclude the extreme scenario in which all T\.ZOs accrete at a rate  $\gtrsim 10^6 M_\odot\,$~yr$^{-1}$. \label{fig:diffuse}}

\end{figure}

The cumulative neutrino flux from all past core collapses constitutes the diffuse supernova neutrino background~\citep[DSNB]{Ando:2023fcc}.
Analogously, the diffuse neutrino flux from all T$\dot{\rm Z}$Os should  permeate the Universe, partially overlapping in energy with the DSNB and with the diffuse neutrino flux from neutron-star mergers, which is expected to be lower than the DSNB~\citep{{Tamborra:2024fcd}}. 

This diffuse T$\dot{\rm Z}$O neutrino flux  is given by
\begin{align}
\Phi_{\rm diffuse} (E_\nu) = \int_0 ^ {z_{\rm max}} \textrm{d}z \frac{\mathcal{R}_{\rm T\dot{Z}O}(z, \dot{M})}{H(z)}\frac{{\rm d}N_\nu}{{\rm d}E_\nu}\bigg|_{E_\nu (1 + z)}\,,
\label{eqn:diffuse}
\end{align}
where $H(z)$ is the Hubble parameter, $\mathcal{R}_{\rm T\dot{Z}O}(z, \dot{M})$ is the cosmic rate of T$\dot{\rm Z}$Os at a given redshift $z$ and for an accretion rate $\dot{M}$, and  ${\rm d}N_\nu/{\rm d}E_\nu$ is the neutrino energy distribution from  a T$\dot{\rm Z}$O (with duration given by  Eq.~\ref{eqn:tbh} and where we consider a constant mass accretion rate such that  $\Delta m = 1\, M_\odot$). Moreover,  we assume that the duration of the neutrino emission from a T$\dot{\rm Z}$O is much shorter than the cosmological timescales. 

We assume that the rate of T$\dot{\rm Z}$Os follows the star formation history, $\dot{\rho}_*$, parametrized as in~\citet{Horiuchi:2008jz}. We take the local rate of T$\dot{\rm Z}$Os to be $\xi_0 = 10^{-4}M_\odot^{-1}$~\citep{Nathaniel:2024mde}. 
Then, the cosmic rate is
\begin{align}
\label{eq:rate}
\mathcal{R}_{\rm T\dot{Z}O}(z, \dot{M}) = \xi_0 \, \, \dot{\rho}_*(z)   \,\, \delta(\dot{M} - \dot{M}_0)\, ,
\end{align}
where $\delta$ denotes the Dirac delta distribution.
We take into account  the contribution from T$\dot{\rm Z}$Os up to redshift $z_{\rm max} =5$, since the contribution from larger redshifts should only matter for neutrinos with  energy below $1$~MeV, whose detection prospects are not optimal. In addition, the equation above assumes that the accretion rate is the same for all T$\dot{\rm Z}$O in the Universe ($\dot{M}_0$).
However, we stress that the rate of T\.ZOs is largely uncertain. For example,  assuming that T\.ZOs are directly linked to gamma-ray burst progenitors, the rate of T\.ZOs can be  inferred from the one of gamma-ray bursts~\citep{Fryer1999, Belczynski2002,ivanova2003}. In this case,  the expected local rate of He stars--neutron star mergers is $\sim 10^{-3}$--$1$~Mpc$^{-3}$~Myr$^{-1}$. The latter is up to two orders of magnitude smaller than the local rate provided in~\cite{Nathaniel:2024mde}, i.e.~$1.6$~Mpc$^{-3}$~Myr$^{-1}$. 

Current searches for a flux of neutrinos of astrophysical origin in the energy range of interest (i.e., $10$--$100$~MeV) focus on electron antineutrino detection through inverse beta decay. Figure~\ref{fig:diffuse} displays the predicted diffuse flux of electron antineutrinos for our selected  accretion rates together with the  DSNB for comparison. The white vertical band marks the  region of interest (ROI) for DSNB searches  between $10$  and  $30$~MeV~\citep{Vitagliano:2019yzm}, and the dashed vertical line indicates the detection threshold for inverse beta decay. Current limits on the DSNB  from Super-Kamiokande~\citep{SK-Gd:2024} already exclude the extreme scenario of all T\.ZOs accreting at a rate of $10^6$ $M_\odot$~yr$^{-1}$ or larger for the  largest T\.ZO local rates, e.g. 1.6 Mpc$^{-3}$ Myr$^{-1}$. Note, that neutrinos from accretion disks, including those from TZOs, could give a relevant contribution to the MeV diffuse flux of astrophysical neutrinos, as also reported in preliminary estimations on the subject~\citep{Schilbach:2018bsg}. However, for most of the predicted merger rates, the T\.ZO contribution to the diffuse neutrino background in the MeV region  would be negligible.

\section{Conclusions}
\label{sec:conclusions}
In this {\it Letter},  we compute the neutrino emission from  Thorne-\.Zytkow Objects (T$\dot{\rm Z}$Os)  and explore their detection prospects.
We distinguish between transient sources (with accretion rates ranging between $10^4 M_\odot$~yr$^{-1}$ and $10^6 M_\odot$~yr$^{-1}$) and steady sources (with  accretion rates as low as $10^{-4}\, M_\odot$ yr$^{-1}$). For transient sources, we rely on a steady accretion disk model, whereas for the steady T\.ZOs,  we consider a steady spherically-symmetric model. In both scenarios, we assume constant accretion.
If hypercritical (super-Eddington) accretion rates are achieved, we find that the high temperature and electron number density in the proximity of the  neutron star lead to a sizable production of thermal neutrinos,  mainly via beta processes and electron-positron pair annihilation.  For the accretion-rate range considered in this work, the expected neutrino flux  peaks between $2$ and $20$~MeV, varying  approximately $9$ orders of magnitude in amplitude. 
 
The duration of the expected neutrino signal could largely vary between a few milliseconds and $\mathcal{O}(10^{11})$~s. Hence, in order to investigate the detection chances of T$\dot{\rm Z}$O neutrinos, we distinguish between ``T$\dot{\rm Z}$O bursts'' (if the accretion rate is $\gtrsim 10^4 M_\odot$~yr$^{-1}$) and ``steady T$\dot{\rm Z}$Os'' (for accretion rates that are $\lesssim 10^4 M_\odot$~yr$^{-1}$).
The  IceCube and Super-Kamiokande neutrino observatories are already sensitive to  T$\dot{\rm Z}$O bursts, even beyond the Small Magellanic Cloud (i.e., $\gtrsim 60$~kpc), if the burst duration should be larger than a few seconds.
Super-Kamiokande  and  the soon-to-be operative JUNO can extend the sensitivity to accretion rates as low as $1 M_\odot$~yr$^{-1}$, if sustained for periods of time ranging from a month to years. In the near future, the Hyper-Kamiokande neutrino observatory will further strengthen these detection prospects. 

Intriguingly, Super-Kamiokande and IceCube could already test the  neutrino emission from existing T$\dot{\rm Z}$O candidates. As for the nearby source, VX Sgr, located at $1.5$--$1.7$~kpc, if this is a T$\dot{\rm Z}$O with an accretion rate  larger than $10^{-3}\, M_\odot$~yr$^{-1}$, targeted  searches should detect neutrinos with large statistics, independent of the uncertainties on the source model. The candidates HV 2112 and HV 11417 are located at about $60$~kpc from Earth. Super-Kamiokande would be able to detect neutrinos from these sources, if they are T$\dot{\rm Z}$Os with an accretion rate larger than $0.5\, M_\odot$~yr$^{-1}$. 

Such encouraging neutrino detection prospects should  motivate the establishment of dedicated searches to test neutrino emission from T$\dot{\rm Z}$Os with existing and upcoming neutrino telescopes. This is especially timely, since neutrinos have the unique potential to assess the nature of  T$\dot{\rm Z}$O candidates. In fact, the electromagnetic signal does not allow us to discriminate whether a neutron star is hosted at the core of such objects, and their gravitational wave emission could only be tested by upcoming gravitational wave detectors.
The diffuse emission of neutrinos coming from all T$\dot{\rm Z}$Os in our Universe falls in the same energy region of the diffuse supernova neutrino background. We show that depending on the cosmic rate of T\.ZOs and their accretion rates, the diffuse flux of neutrinos from T\.ZOs could be comparable to the diffuse supernova neutrino background. In fact, in case of a positive detection of a diffuse flux of astrophysical neutrinos in this energy range, it might be nontrivial to asses whether they originated from past core-collapse supernova or T\.ZOs. However, current limits on the DSNB already exclude extreme scenarios for T\.ZO accretion and formation rates. 

This work highlights the relevance of neutrinos as key messengers to ascertain the existence of  T$\dot{\rm Z}$Os in our Galaxy, and in certain cases even in the Local Group. While we employ a simplified model to compute the neutrino emission from these sources, our work highlights  the importance of carrying out follow-up work on the  neutrino signals from  T$\dot{\rm Z}$Os and lays down  specific detection strategies  to test the nature of T$\dot{\rm Z}$O candidates and their properties. All in all, complementing  ongoing observational efforts across wavelengths, neutrino searches would be of pivotal importance  to probe this population of exotic stellar objects.

\section*{Acknowledgments}
We thank the anonymous referee for valuable suggestions.
This project has received support from the Villum Foundation (Project No.~13164), the Carlsberg Foundation (CF18-0183), and the Deutsche Forschungsgemeinschaft through Sonderforschungbereich SFB 1258 ``Neutrinos and Dark Matter in Astro- and Particle Physics'' (NDM). Part of this work was performed at Aspen Center for Physics, which is supported by National Science Foundation grant PHY-2210452. The Tycho supercomputer hosted at the SCIENCE HPC Center at the University of Copenhagen was used for supporting the numerical simulations presented in this work.

\newpage
\appendix
\section{Source modelling and thermal neutrino production}
\label{sec:app-source-nu}

In this appendix, we introduce the steady accretion disk and spherically-symmetric models that we  employ to compute the neutrino emission from T$\dot{\rm Z}$Os. Moreover, we outline the main reaction channels adopted to compute  neutrino production, and the approach adopted to compute flavor conversion.


\subsection{Steady accretion disk T$\dot{\rm Z}$O model}
\label{sec:disk}

We closely follow \cite{DiMatteo:2002iex} to model the neutrino emission from the accretion disk forming around the compact object in the T\.ZO  core  for accretion rates larger than $10^4 M_\odot$~yr$^{-1}$. 
To this purpose, we solve the continuity equation,
\begin{align}
    \dot{M} = 4\pi R \rho H v\, ,
\end{align}
and the energy-conservation equation,
\begin{equation}
    \frac{3G M_{\rm CO}\dot{M}}{8\pi R^3} = q_{\nu} + q_{\rm adv}\, . 
\end{equation}
Here, $R$ is the radial distance in cilindrical coordinates, $H$ is the vertical scale height, $v$ is the radial velocity of the gas, $M_{\rm CO}$ is the mass of the compact object, and $q_\nu$ ad $q_{\rm adv}$ are the cooling rates due to neutrino losses and advection respectively. We also adopt the following equation of state:
\begin{align}
    P = \rho \frac{k_B}{m_p} T \frac{1 + 3X_{\rm nuc}}{4} +\frac{11}{12}aT^4 + K(\rho Y_e)^{4/3}\, ,
\end{align}
where $k_B$ is the Boltzmann constant, $K = 1.24\times 10^{15}$, and $X_{\rm nuc}$ is the mass fraction of nucleons. This approach is based on Newtonian dynamics and is overall consistent with other steady models, such as the one proposed in \cite{Popham:1998ab}.

We focus on the scenario in which the central compact object is a neutron star and follow the approach presented in \cite{Zhang:2007ds} to describe the inner part of the disk with a self-similar structure. This solution is applied to radius smaller than  $r_{\rm in} \sim 20$~km~\citep{Popham:1998ab}.
As a result, the disk is advection-dominated in the outer region whereas neutrino cooling becomes relevant in the innermost regions.
In particular,  due to the large temperature and density in the surrounding of the neutron star, helium  photodissociates and beta processes can take place.

We assume a neutron star with mass equal to $1.4 M_\odot$ and  radius of $r_{\rm NS} = 10$~km. We assume that the disk extends up to $100\ r_{\rm NS}$ and the electron fraction is constant and equal to $Y_e = 0.5$. In the top panels of Fig.~\ref{fig:elr}, we show the density and temperature profiles corresponding to accretion disks where the accretion rate is $10^4 \, M_\odot$~yr$^{-1}$ and $10^6 \, M_\odot$~yr$^{-1}$. 

\subsection{Steady and spherically-symmetric T$\dot{\rm Z}$O model}
\label{sec:sphere}
In order to model the neutrino emission from spherically symmetric T$\dot{\rm Z}$Os with accretion rate smaller than $10^4 M_\odot$~yr$^{-1}$, we follow the analytical approach presented in~\cite{Chevalier:1989}.
To this purpose, we consider a steady and spherically-symmetric inflow of shocked gas surrounding the
neutron star. This approach also assumes that the fluid velocity is subsonic and  the postshock flow is adiabatic. The first assumption is not valid  in the vicinity of the shock wave; whereas the second assumption relies on the fact that neutrino energy losses are not negligible in the proximity of the neutron star. Within this framework,  we derive the density and temperature profiles of the T$\dot{\rm Z}$O envelope. Moreover, for simplicity, we consider a constant electron fraction, $Y_e = 0.5$.

As for the  neutron-star properties, we assume  a mass of $M_{\rm NS} = 1.4 \, {M}_\odot$ and a radius $r_{\rm NS} = 10$~km, with  constant density and temperature within $r_{\rm NS}$, equal to the temperature at its surface. Note that the density and temperature profiles in the envelope are negligibly affected by such assumptions. 
The baryon density profile is
\begin{align}
    \rho(r) = \begin{cases}
        \dfrac{3M_\textrm{NS}}{4\pi r_\textrm{NS}^3} & r \leq r_\textrm{NS}\\[1em]
        \rho_{\textrm{sh}} \left(\dfrac{r}{r_\textrm{sh}}\right)^{-3} & r_\textrm{NS} < r \leq r_\textrm{sh} \\[1em]
        \dfrac{\dot{M}}{4\pi r_\textrm{sh}^2 v_\textrm{ff}(r)} & r > r_\textrm{sh}
    \end{cases} \, \quad ,
\end{align}
where we define the characteristic density ($\rho_{\textrm{sh}}$) and the free-fall velocity ($v_{\rm{ff}}$) as
\begin{align}
    \rho_{\textrm{sh}} = 7 \frac{\dot{M}}{4\pi r_\textrm{sh}^2 v_\textrm{ff}(r_{\textrm{sh}})}\,  \quad  \textrm{and } \quad v_\textrm{ff} (r)= \sqrt{\frac{2 G {\rm M}_\textrm{NS}}{r}}\, .
\end{align}
Here, $G$ is the gravitational constant, $\dot{M}$ is the accretion rate, $r_{\textrm{sh}}$ indicates the location of the accretion-shock radius~\citep{Chevalier:1989}:
\begin{align}
    r_{\textrm{sh}} = 3.6 \times 10^8 \left(\frac{\dot{M}}{M_\odot \,\,\textrm{yr}^{-1}}\right)^{-4/10} \left(\frac{r_{\textrm{NS}}}{10^6 \, \textrm{cm}}\right)^{8/5}\left(\frac{M_\textrm{NS}}{1.4 \, M_\odot }\right)^{-1/15}\, \text{cm}\,.
\end{align}

\begin{figure*}
    \centering
    \includegraphics[width=\linewidth]{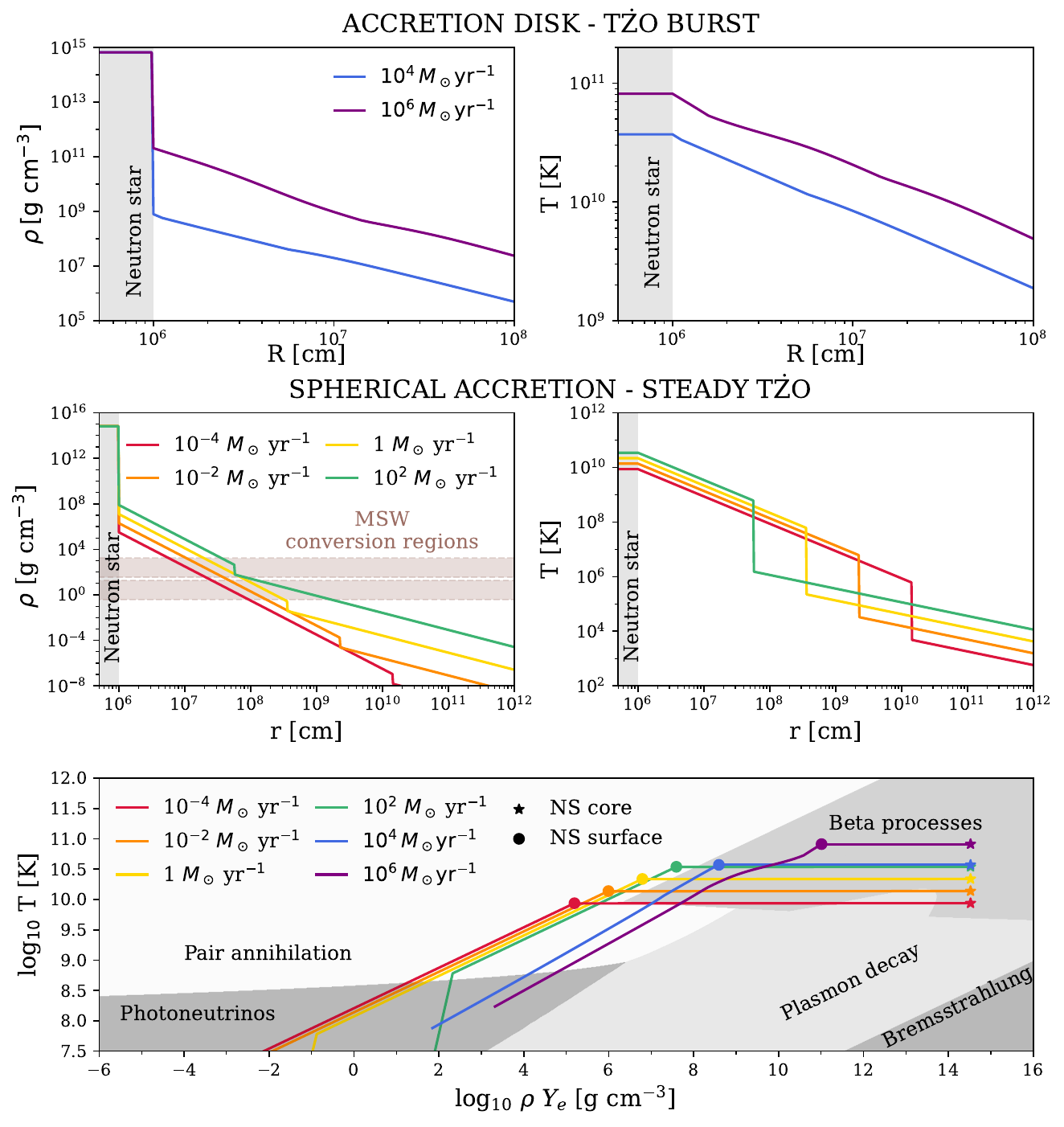}
    \caption{{\it Top panels:} Radial evolution of the baryon density on the left and temperature on the right, for different accretion rates for a steady disk. {\it Middle panels: }Radial evolution of the baryon density on the left and temperature on the right, for different accretion rates in the steady spherically symmetric scenario. The brown horizontal bands in the top left panel highlight the regions where MSW flavor conversion is expected to take place for the solar and atmospheric neutrino mass differences and $E_\nu \in [2, 100]$~MeV. {\it Bottom panel:} Dominant mechanisms for thermal neutrino production in the plane spanned by the electron density and temperature of the medium. The shaded regions indicate the  dominant processes: beta processes, electron-positron pair annihilation, plasmon decay, photoneutrino production, and bremsstrahlung~\citep{Dicus:1972yr,Dutta:2003ny, Braaten:1993jw,Guo:2016vls,Haft:1993jt,Qian:1996xt}. The colored lines correspond to the density and temperature profiles for different accretion rates from $10^{-4}$ to 10$^6 \, M_\odot$~yr$^{-1}$. The star-shaped markers indicate the characteristic temperature and density of the neutron star (assumed to be constant for simplicity);  the point-like markers signal the thermodynamical quantities right outside of the neutron star (assuming a sharp boundary).  \label{fig:elr}}
\end{figure*}

Similarly, the temperature profile scales as
\begin{align}
    {T}(r) = \begin{cases}
        {T}_\textrm{NS} & r \leq r_\textrm{NS}\\[1em]
        {T}_\textrm{sh} \left(\dfrac{r}{r_\textrm{sh}}\right)^{-1} & r_\textrm{NS}< r \leq r_\textrm{sh} \\[1em]
        {T}_\textrm{sh}\left(\frac{\rho(r)}{\rho_\textrm{sh}}\right)^{1/3} & r > r_\textrm{sh}
    \end{cases} \, \quad ,
\end{align}
where the temperature at the shock front ($T_\textrm{sh}$) and at the neutron-star surface ($T_\textrm{NS}$) are given by
\begin{align}
    {T}_\textrm{sh} = \left(\frac{6}{7} \frac{\dot{M} v_\textrm{ff}(r_\textrm{sh})}{4\pi r^2_\textrm{sh} a}\right)^{1/4} \quad \text{and} \quad {T}_\textrm{NS} = {T}_\textrm{sh} \frac{r_\textrm{sh}}{r_\textrm{NS}}\, ,
\end{align}
with $a = 4\sigma/c$,  $\sigma$ is the Stefan–Boltzmann constant, and $c$ is the speed of light.
Given the wide range of accretion rates explored in this paper, the temperatures at the surface of the neutron star can vary between $10^{10}$ and $10^{11}$~K. As for the baryon density at the innermost layer of the envelope, it can vary between  \mbox{$10^4$~g~cm$^{-3}$} and $10^9$~g~cm$^{-3}$. The resulting temperature and density profiles for different mass accretion rates are plotted in the middle panels of Fig.~\ref{fig:elr}.

\subsection{Neutrino emission rates}
In regions with $T \gtrsim 10^{10}$~K , alpha particles can be photodissociated. Since the medium is dense  in free neutrons and protons, beta  processes ($ p + e^- \leftrightharpoons\nu_e + n  $ and $n + e^+ \leftrightharpoons  \bar{\nu}_e + p $) can lead to  emission and absorption of electron neutrinos and antineutrinos. Photodissociation is relevant at large baryon densities, $\rho \gtrsim 10^8$ g cm$^{-3}$, as it is the case for the innermost regions of accretion disks. As a consequence, beta processes dominate the neutrino emission in the surroundings of the neutron star for  disks accreting at rates $\dot{M}\leq 10^4 M_\odot$~yr$^{-1}$. For simplicity, we assume that these reactions emit neutrinos following a thermal distribution~\citep{Patton:2017neq} and take $Q = m_n - m_p \simeq 0$. Note also that we do not account for the change in the electron fraction induced by the beta processes  and assume  $Y_e = 0.5$ instead. Under these assumptions, the estimated flux of electron neutrinos and antineutrinos is equal.

The neutrino emission rate through pair annihilation ($e^+ + e^- \to \nu + \bar{\nu}$, \citealt{Dicus:1972yr}) and photoneutrino production ($\gamma + e^- \to e^- + \nu + \bar{\nu}$, \citealt{Dutta:2003ny}) also grows with temperature. In addition, the pair annihilation rate  presents a mild growth as the electron density increases. As shown in the bottom panel of Fig.~\ref{fig:elr}, in the surroundings of a spherically-accreting neutron star (for $10^{-4}M_\odot$~yr$^{-1} \leq \dot{M} < 10^4 M_\odot$~yr$^{-1}$), neutrino emission is  dominated by  pair annihilation in the  vicinity of the neutron star and photoneutrino production at smaller densities.

In the neutron star, the main neutrino production channel is plasmon decay ($\gamma^* \to \nu + \bar{\nu}$, \citealt{Braaten:1993jw}). However, due to the very high densities [$\mathcal{O}$(10$^{14})$~g~cm$^{-3}$], we find that absorption is so large that the neutron star can be safely treated as opaque to neutrinos.

In the absence of flavor conversion, the neutrino flux at Earth for a given neutrino of flavor $\alpha$ from the T\.ZO accretion disk (cf.~Sec.~\ref{sec:disk}) is
\begin{align}
    \frac{{\rm d^2}\Phi_{{\nu}_\alpha,\, 0}}{{\rm d}E_\nu{\rm d}t} = \int {\rm d}R\, \frac{R}{2 D^2} H(R) \,\, \left[1 +\sqrt{1 - \frac{R_{\rm dec}^2}{R^2 + H(R)^2}}\,\,\right] \, \,\mathcal{R}_{\nu_\alpha}(\tilde{E}(r), \rho(r), T(r))\, .
\end{align}

It depends on the neutrino emission rate for given  temperature and baryon density. The energy at which neutrinos are produced and  detected  are  $\tilde{E}$ and $E_\nu$, respectively, with $\tilde{E} = E_\nu /\sqrt{1 - 2G{M}_{\rm NS}/r}$; the redshift correction on the neutrino energy is due to the fact that the spacetime is curved  in the proximity of the neutron star. We have also included the geometric factor $1/2 \left[1 + \sqrt{1-(r_{\rm NS}^2/(H(R)^2 + R^2)}\right]$, which accounts for the fact that neutrino absorption is negligible in the envelope outside the neutron star, but the neutron star is completely opaque to neutrinos.
Notice also that only the emission from radii larger than the decoupling radius, $R_{\rm dec}$, is taken into account. Although we have considered an outer radius of 1000 km, since the neutrino emission originates in the several tens of kilometers surrounding the neutron star, the event rate expected for a disk of 50~km is less than 10\% smaller.

Analogously, for a spherically symmetric emission (cf.~Sec.~\ref{sec:sphere}), the neutrino flux at Earth  is defined as
\begin{align}
    \frac{{\rm d^2}\Phi_{{\nu}_\alpha,\, 0}}{{\rm d}E_\nu{\rm d}t} = \int {\rm d}r\, \frac{r^2}{2 D^2}\,\, \left[1 +\sqrt{1 - \frac{r_{\rm NS}^2}{r^2}}\,\,\right] \, \,\mathcal{R}_{\nu_\alpha}(\tilde{E}(r), \rho(r), T(r))\, .
\end{align}

In this case, all regions outside the neutron star are optically thin to neutrinos and the corresponding geometric factor is $1/2 \left[1 + \sqrt{1-(r_{\rm NS}/r)^2}\right]$.

\subsection{Neutrino flavor conversion}

Lepton number conservation ensures that the same amount of neutrinos and antineutrinos of each flavor are produced. Electron flavor neutrinos are produced via charged and neutral weak interactions, whereas non-electron  neutrinos are produced only through neutral current interactions. Hence, the fraction of electron flavor neutrino-antineutrino pairs is larger than the one of each non-electron flavor neutrino-antineutrino pair. 

Neutrinos are mainly produced in a region extending for several tens of km outside the neutron star. When travelling outwards, they experience Mikheyev–Smirnov–Wolfenstein (MSW) resonant flavor conversion~\citep{Wolfenstein:1977ue,Mikheyev:1985zog} when they cross the baryon density  at $\sim 10^2$--$10^3$~g~cm$^{-3}$ and $\sim 1$--$10$~g~cm$^{-3}$, as illustrated in the top left panel of Fig.~\ref{fig:elr}. We have tested that, due to the source properties,   neutrino flavor evolution in the steady T$\dot{\rm Z}$O envelope is adiabatic. Then, taking into account the loss of coherence for neutrinos en route to  Earth, as illustrated in~\cite{Dighe:1999bi}, we  obtain the  (anti)neutrino fluxes at Earth. In this work, the three-flavor neutrino oscillation parameters are set as in~\cite{deSalas:2020pgw}. For the case of a transient T\.ZO, we do not model the density profile beyond the disk and hence, we can not test if flavour evolution is adiabatic. However, for simplicity, we consider that the same description of flavour conversions applies.

\section{Neutrino event rate in existing and upcoming neutrino telescopes}
\label{sec:app-nu-obs}

In this appendix, we discuss how to compute the event rate of neutrinos from T$\dot{\rm Z}$Os in existing neutrino telescopes, such as IceCube and Super-Kamiokande, as well as in the upcoming  Hyper-Kamiokande and JUNO.

\begin{figure}
    \centering
    \includegraphics[width= \linewidth]{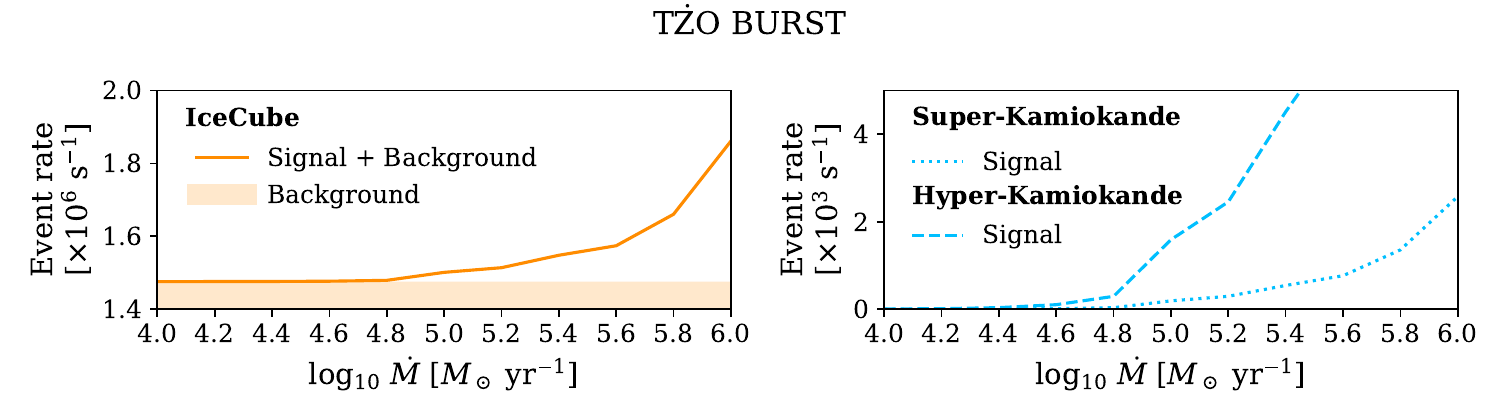}
    \includegraphics[width=\linewidth]{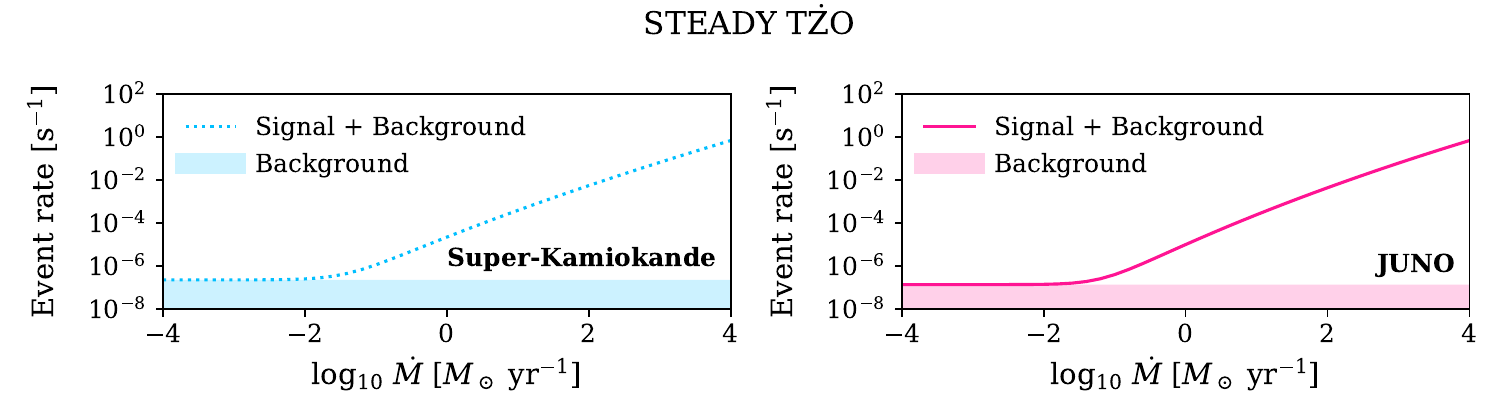}
    \caption{{\it Top panels:} Expected event rate for a T$\dot{\rm Z}$O burst at $5$~kpc from Earth for IceCube (left panel), Super-Kamiokande (dotted line, right panel), and Hyper-Kamiokande (dashed line,  right panel). The IceCube background rate is shown as a shaded band. For Super-Kamiokande (Hyper-Kamiokande), the background rate is $10^{-2}$~s$^{-1}$ ($0.1$~s$^{-1}$), and thus  not displayed.
    {\it Bottom panels:} Expected event rate for a steady T$\dot{\rm Z}$Os at $5$~kpc from Earth for Super-Kamiokande (left panel) and JUNO (right panel). The background rates are displayed as shaded bands. The total and background-only event rates for Hyper-Kamiokande are analogous to the ones shown for Super-Kamiokande, after accounting for the difference in fiducial volume of approximately one order of magnitude (and therefore not shown here).}
    \label{fig:event-rates}
\end{figure}

\subsection{IceCube Neutrino Observatory}
The IceCube Neutrino Observatory is a water-Cherenkov detector located at the South Pole and consisting of $5160$ digital optical modules (DOMs). IceCube is expected to be sensitive to  neutrinos from T$\dot{\rm Z}$Os via (i)~the inverse beta decay channel ($\bar{\nu}_e + p \to e^+ + n$); (ii) the elastic scattering of neutrinos of all flavors on electrons ($\nu_\alpha + e^-\to \nu_\alpha + e^-$ and $\bar{\nu}_\alpha + e^-\to \bar{\nu}_\alpha + e^-$, with $\alpha = $ e$,\, \mu,\,\tau$); (iii)~the quasi-elastic scattering of electron neutrinos on oxygen ($\nu_e + ^{16}$O$ \to e^ - + X$ and $\bar{\nu}_e + ^{16}$O$ \to e^ + + X$, where $X$ denotes a different final state nucleus) by measuring the Cherenkov photons radiated by the final-state electrons and positrons. 

Following \cite{IceCube:2011cwc}, we compute the hit count per DOM for each reaction channel as 
\begin{align}
    r_{\rm IC} = n_{\rm target} \int {d}E_e\int {d}E_\nu\,\frac{{d}\sigma(E_\nu, E_e)}{{d}E_e} \,\, N_\gamma(E_e) \,\, V_\gamma^{\rm eff} \,\, \Phi(E_\nu)\, ,
\end{align}
where $n_{\rm target}$ is the number density of targets for a density of $0.92$~g~cm$^{-3}$. The effective volume for a single photon is $V_\gamma^{\rm eff} = 0.163$ and the energy-dependent number of radiated photons as a function of the electron or positron kinetic energy  ($E_e$) is $N_\gamma(E_e) \simeq 178\, E_e /{\rm MeV}$. Finally, ${d}\sigma/{d}E_e$ denotes the cross-section for  the detection channels listed above. Such cross-sections have been implemented following~\cite{Strumia:2003zx,Ricciardi:2022pru,Marciano:2003eq,Kolbe:2002gk}.

Defining $r$ as the sum of the hit count per DOM for all reaction channels, the expected event rate is 
\begin{align}
    R_{\rm IC} = N_{\rm DOM}\,\, r \,\, \frac{0.87}{1 + r\,\, \tau}  \, ,
\end{align}
where $\tau \simeq 250$~s is the deadtime and $N_{\rm DOM} = 5160$ is the number of DOMs. 
The signal event rate is to be compared with the rate of background events and shot-noise, that is $1.5\times 10^6$~s$^{-1}$~\citep{IceCube:2011cwc}, see the top left panel of Fig.~\ref{fig:event-rates}. The large background rate makes IceCube unsuitable for the search of steady MeV neutrino sources. Besides that, IceCube can measure the total rate but can not recover accurate information on the energy spectrum.

\subsection{Super-Kamiokande and Hyper-Kamiokande}

The water-Cherenkov detector Super-Kamiokande is sensitive to T$\dot{\rm Z}$O neutrinos via inverse beta decay, elastic scattering on electrons, and quasi-elastic scattering on oxygen, like IceCube.
The event rate for each detection channel is 
\begin{align}
    R_{\rm SK}(t) = n_{\rm targets} \,\, \varepsilon \, \int {d}E_e \int {d}E_\nu \frac{{d}\sigma(E_\nu, E_e)}{{d}E_e} \,\,\Phi(E_\nu)\, ,
\end{align}
and we consider a fiducial volume of $22.5$~kT of water.
Neutrino searches from T$\dot{\rm Z}$O bursts are basically background free (the background rate is $0.012$~s$^{-1}$, generally much smaller than the expected signal event rate, c.f.~Table~\ref{tab:evrates}).
We assume  perfect detection efficiency ($\varepsilon = 1$) for all  detection channels. 

For steady T$\dot{\rm Z}$Os, the detection efficiency is assumed to be $\varepsilon = 0.55$ (resulting from a $75\%$ neutron-capture efficiency for $0.03\%$ of Gadolinium-loaded water and  $74\%$ selection efficiency,~\citealt{Super-Kamiokande:2024kcb}).  The detection window is limited to neutrino energies between $9.3$ and $31.3$~MeV. For steady T$\dot{\rm Z}$Os, we only consider inverse beta decay events.  These searches need to account for the background events from reactor antineutrinos, atmospheric neutrinos, as well as spallation backgrounds, and the DSNB. These backgrounds are included following~\cite{Martinez-Mirave:2024zck} and references therein. 
In the future, Hyper-Kamiokande~\citep{Hyper-Kamiokande:2018ofw} will further improve the detection prospects explored for Super-Kamiokande, due to its larger fiducial volume ($187$~kT of water). We assume a simple scaling in detector size from Super-Kamiokande, following~\cite{Martinez-Mirave:2024zck}. 
The expected event rates as functions of the accretion rates for T$\dot{\rm Z}$O bursts and  steady T$\dot{\rm Z}$Os are shown in Fig.~\ref{fig:event-rates} for both Super-Kamiokande and Hyper-Kamiokande. 

\subsection{JUNO}
The liquid-scintillator neutrino detector JUNO is expected to be operative within the next year. While the event statistics expected in Super-Kamiokande and IceCube from T$\dot{\rm Z}$O bursts is larger than the one  achievable in JUNO, the latter would be suitable for searches of neutrinos from steady T$\dot{\rm Z}$Os. 
JUNO is expected to be sensitive to electron antineutrinos via the inverse beta decay channel, with an event rate  given by:
\begin{align}
    R_{\rm JUNO} = n_{\rm targets} \,\, \varepsilon \,\, \int {d}E_e \int
{d}E_\nu\, \frac{{d}\sigma_{\rm IBD}(E_\nu, E_e)}{{d}E_e}  \,\, \Phi_{\bar{\nu}_e} (E_\nu)\,.
\end{align}
Here, the number of proton tagets is $n_{\rm targets} =1.22\times 10^{33}$, equivalent to $17$~kT,  and the detection efficiency is $\varepsilon = 0.8$~\citep{JUNO:2022lpc}. The event rate depends on the electron antineutrino flux ($\Phi_{\bar{\nu}_e}$) and the inverse beta decay cross section ($\sigma_{\rm IBD}$).
The main backgrounds expected in JUNO are due to reactor antineutrinos, atmospheric neutrinos, spallation backgrounds, and the DSNB~\citep{JUNO:2022lpc}. We take such backgrounds into account following~\cite{Martinez-Mirave:2024zck}. 
The event rate expected in JUNO for steady T$\dot{\rm Z}$Os is shown in Fig.~\ref{fig:event-rates} (bottom right panel) with the related background rates.

\bibliography{bibliography}
\bibliographystyle{aasjournal} 

\end{document}